\documentclass[12pt]{iopart}
\usepackage{graphicx}
\pdfoutput=1

\begin{document}

\title{Dissipation characteristics of quantized spin waves in nano-scaled magnetic ring structures}

\author{H.~Schultheiss, C.W.~Sandweg, B.~Obry, S.~Hermsd\"orfer, S.~Sch\"afer, B.~Leven, and B.~Hillebrands}

\address{Fachbereich Physik and Forschungsschwerpunkt MINAS,
Technische Universit\"at Kaiserslautern, Erwin-Schr\"odinger-Stra{\ss}e 56, 67663 Kaiserslautern, Germany}
\ead{helmut@physik.uni-kl.de}
\begin{abstract}
The spatial profiles and the dissipation characteristics of spin-wave quasi-eigenmodes are investigated in small
magnetic Ni$_{81}$Fe$_{19}$ ring structures using Brillouin light scattering microscopy. It is found, that the
decay constant of a mode decreases with increasing mode frequency. Indications for a contribution of three-magnon
processes to the dissipation of higher-order spin-wave quasi-eigenmodes are found.
\end{abstract}

\maketitle

\section{Introduction}
The investigation of the magnetic excitation spectrum of magnetic nano-structures is of large importance for the
fundamental understanding of the magnetization dynamics as well as for technological applications. The demands
concerning speed and reliability for future logic and data storage devices are pushing the operating frequency
towards the GHz-regime -- the typical time-scale of spin waves, the fundamental excitations of the magnetization.
Since the discovery of the spin-wave quantization effect in patterned magnetic thin films
\cite{Mathieu:1998p628,Jorzick:2002p81} many works focused on the research of eigen-excitations in magnetic media
with a reduced dimensionality. The amplitude profile, the discrete frequencies and the quantization conditions of
spin-wave eigenmodes in stripes
\cite{Mathieu:1998p628,Jorzick:2002p81,PhysRevLett.91.137204,Demokritov:2001p2297}, rectangles
\cite{Demokritov:2001p2297,Gubbiotti:2004p160}, disks and ellipses \cite{Gubbiotti:2005p109,Gubbiotti:2003p7} are
now well understood and allow nowadays even for an engineering of the spin-wave eigenmode spectrum in such a
small magnetic structure taking into account issues of geometric shape, multilayer stacks, and even magnetic
domain structure.

The magnetization dynamics of rings magnetized in the onion state shows a large richness in its eigenmode system.
Quantization of spin waves takes place not only due to the confinement in radial and azimuthal direction in the
so-called equatorial regions at \mbox{90$^{\circ}$} and \mbox{270$^{\circ}$} (see inset in Fig.\,\ref{Fig_1} for
the definition of the angular positions) but also in spin-wave wells in the so-called pole regions at
\mbox{0$^{\circ}$} and \mbox{180$^{\circ}$} created by the inhomogeneity of the total internal field
\cite{schultheiss:047204}. The spatial profiles of these eigenmodes have been widely discussed in the last years
by several research groups
\cite{schultheiss:047204,Neudecker:2006p93,Gubbiotti:2006p138,Zhu:2006p105,Giesen:2005p58,Giesen:2005p146} and it
is well known, that low-frequency spin waves are located at the poles of the onion state whereas in the
equatorial regions confined spin waves with higher frequencies are present with nodes either in azimuthal or
radial direction, depending on the excitation.

In the real world modes are not true eigenmodes in a sense, that they are not totally decoupled from the environment
and from each other. They are weakly coupled to the environment, i.e. to the lattice, and thus they are weakly
lossy. So far the problem of dissipation has been rarely addressed. In \cite{schultheiss:047204} the issue of
finite coherence length of spin waves in a ring structure is discussed. For a deeper understanding it is of
central interest to study the mode dynamics, in particular how fast a mode is relaxing after resonant excitation
with a short microwave pulse and to identify the channels of dissipation. Since the coupling is weak and can be
discussed in a perturbation picture, we address the mode spectrum in the following as the ``quasi-eigenmode''
spectrum.

\section{Experiment}
\begin{figure}
\includegraphics[width=0.5\columnwidth]{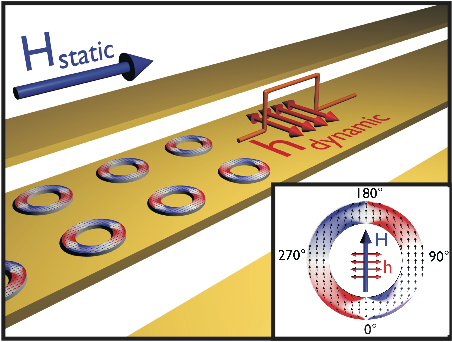}
\caption{\label{Fig_1}%
(color online) The magnetic ring structures with a thickness of \mbox{30\,nm}, an outer
diameter of \mbox{2\,$\mu$m} and a ring width of \mbox{400\,nm} are placed on top of a coplanar waveguide made
out of Au with a central strip line width of \mbox{20\,$\mu$m} and thickness of \mbox{200\,nm}. A static field
H$_{static}$ of \mbox{330\,Oe} is applied in the sample plane and perpendicular to the strip line. The inset
shows a micromagnetic simulation of the so called onion state of the magnetization. For the particular geometry
and applied field transverse domain walls are present at \mbox{0$^{\circ}$} and \mbox{180$^{\circ}$} leading to a
strong reduction of the total internal field. In both regions, the poles at \mbox{0$^{\circ}$} and
\mbox{180$^{\circ}$}, and the equator at \mbox{90$^{\circ}$} and \mbox{270$^{\circ}$}, the coupling of the
dynamic magnetic field h$_{dynamic}$ caused by a microwave current flowing through the coplanar waveguide is most
efficient. }
\end{figure}

Here we report on the investigation of the lifetime of quantized spin waves in a small magnetic ring structure
magnetized in the so-called onion state using time-resolved Brillouin light scattering microscopy. For this
purpose, permalloy (Ni$_{81}$Fe$_{19}$) rings with an outer dia\-me\-ter of \mbox{$D = $2\,$\mu$m}, a ring width
of $w=400$\,nm, and a thickness of \mbox{30\,nm} were prepared by a combination of molecular beam epitaxy in an
UHV system and electron beam lithography, utilizing a lift-off technique. The ring structures are placed on top
of a coplanar waveguide as shown schematically in Fig.\,\ref{Fig_1}. The waveguide is made out of a 200\,nm Au
layer and the central strip line has a width of \mbox{20\,$\mu$m}, small enough to ensure high excitation fields
in the region where the ring structures are placed. The ring to ring spacing was chosen twice the ring diameter
to avoid magneto-static and dynamic coupling between neighbouring rings. The geometry of the dynamic and static
magnetic fields used in these experiments is illustrated in Fig.\,\ref{Fig_1}: a static magnetic field
H$_{\mathrm{static}}$ of \mbox{330\,Oe} is applied along the direction of the central strip line to induce the
onion state of the magnetization. In this case the magnetization is aligned mainly parallel to the axis of the
central strip line of the coplanar waveguide and, therefore, the efficiency of the excitation with the dynamic
magnetic field h$_{\mathrm{dynamic}}$ is optimal.

\begin{figure}
\includegraphics[width=0.5\columnwidth]{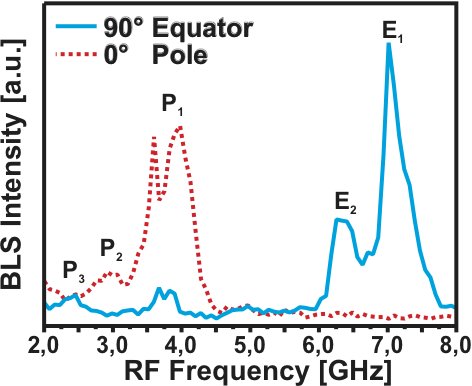}
\caption{\label{Fig_2}%
(color online) Spin-wave amplitudes measured with BLS microscopy under the application of
a dynamic magnetic field in the plane of the sample caused by a microwave current flowing through the central
conductor of the coplanar waveguide. The spectra were acquired in the pole region at \mbox{0$^{\circ}$}, red
dotted line, and in the equatorial region at \mbox{90$^{\circ}$}, blue solid line. Five resonances are found,
three (\mbox{$P_{3}$=2.5\,GHz}, \mbox{$P_{2}$=2.9\,GHz}, \mbox{$P_{1}$=3.9\,GHz}) located at the pole and two
(\mbox{$E_{1}$=6.6\,GHz}, \mbox{$E_{2}$=7.3\,GHz}) at the equator. }
\end{figure}

In order to determine the quasi-eigenmode frequencies in the pole and equatorial regions of the onion state we
performed FMR-BLS measurements at \mbox{0$^{\circ}$} and \mbox{90$^{\circ}$}. A microwave current with
frequencies between 2\,GHz and 8\,GHz was sent through the coplanar waveguide to excite the magnetization. The
precession amplitude is measured locally by means of BLS microscopy with a spatial resolution of \mbox{250\,nm}.
A detailed explanation of this technique can be found in \cite{Perzlmaier:2005p68,Demidov:2004p166}. The results
are shown in Fig.\,\ref{Fig_2} where the BLS intensity is plotted as a function of the applied microwave
frequency for the pole position at \mbox{0$^{\circ}$} (red, dotted line) and the equator position at
\mbox{90$^{\circ}$} (blue, solid line). Two very strong resonances can be observed at \mbox{90$^{\circ}$}:
\mbox{$E_{1}$=7.3\,GHz} and \mbox{$E_{2}$=6.6\,GHz}. Two weaker resonances \mbox{$P_{3}$=2.5\,GHz} and
\mbox{$P_{1}$=3.9\,GHz} with their origin in the pole region, as dicussed in the text below, contribute to the
BLS intensity measured at \mbox{90$^{\circ}$}. The measurement at \mbox{0$^{\circ}$} reveals again $P_{1}$ and in
addition a resonance $P_{2}$ at \mbox{2.9\,GHz}.

\begin{figure}
\includegraphics[width=0.5\columnwidth]{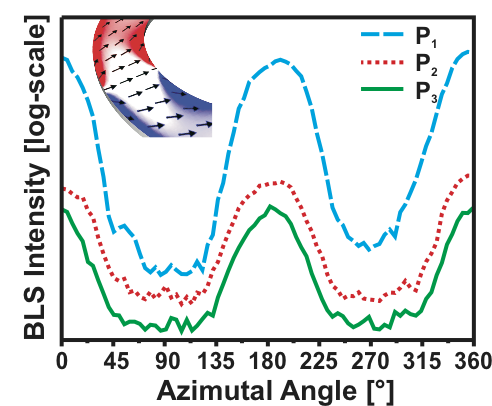}
\caption{\label{Fig_3}%
(color online) Spatial distribution of the spin waves located near the pole regions of the magnetic ring
structure. The spin-wave amplitudes under excitation with the frequencies $P_{1}$, $P_{2}$, and $P_{3}$ of the
quasi-eigenmodes found in Fig.\,\ref{Fig_2} are plotted as a function of the position. For clarity the spectra
are vertically shifted against each other. A strong confinement of the low frequency modes in the poles of a ring
magnetized in the onion state is clearly visible. }
\end{figure}

\begin{figure}
\includegraphics[width=0.5\columnwidth]{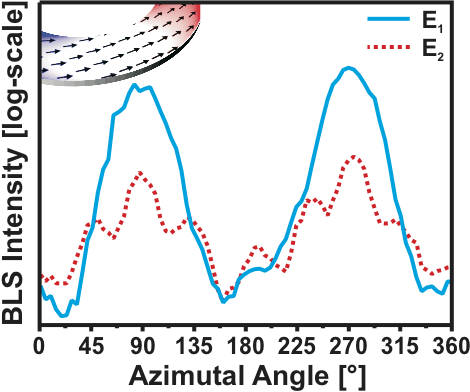}
\caption{\label{Fig_4}%
(color online) Spatial distribution of the spin waves located near the equator regions of the magnetic ring
structure under excitation with the frequencies $E_{1}$ and $E_{2}$ of the quasi-eigenmodes found in
Fig.\,\ref{Fig_2}. The quasi-eigenmode with the highest frequency \mbox{$E_{1}$=7.3\,GHz} (blue solid line) has
its maximum amplitude at \mbox{90$^{\circ}$}. The next quasi-eigenmode \mbox{$E_{2}$=6.6\,GHz} at the equator
shows two nodes in azimuthal direction. }
\end{figure}

For a better understanding of the nature of all the observed resonances we performed BLS measurements as a
function of the azimuthal angle while exciting the magnetization with the resonance frequencies shown in
Fig.\,\ref{Fig_2}. The spatial distributions of the low frequency spin-wave modes $P_{1}$, $P_{2}$, and $P_{3}$
are displayed in Fig.\,\ref{Fig_3} where the BLS intensity is plotted as a function of the azimuthal angle.  All
three modes $P_{1}$, $P_{2}$, and $P_{3}$, observed in Fig.\,\ref{Fig_2}, are strongly confined in the poles of
the onion state, as expected from previous works \cite{schultheiss:047204,Neudecker:2006p93,Gubbiotti:2006p138}. The modes $E_{1}$
and $E_{2}$ with higher frequencies are located in the equatorial region at \mbox{90$^{\circ}$} and
\mbox{270$^{\circ}$} as can be seen in Fig.\,\ref{Fig_4}. The quasi-eigenmode \mbox{E$_1$ at 7.3\,GHz} shows one
maximum and the quasi-eigenmode \mbox{E$_2$ at 6.6\,GHz} shows three maxima in each half of the ring. The fact that higher
azimuthal quasi-eigenmodes in rings and disks have lower frequencies due to their backward volume mode character
is well known and was reported previously \cite{Neudecker:2006p93,Buess:2005p124}.

After the determination of the spin-wave frequencies, which can be resonantly excited with an in-plane magnetic
microwave field, and the positions of maximum spin-wave amplitudes, the investigation of the temporal evolution
of the observed quasi-eigenmodes has been addressed. Therefore, the sample is now excited with microwave pulses
with a duration of \mbox{50\,ns} and a repetition period of \mbox{128\,ns}. The rising edge of the pulse starts a
counter with a time resolution of \mbox{250\,ps}. From this time on, the arrival time of each inelastically
scattered photon is acquired together with its frequency -- similar to the experiment described in
\cite{Demidov:2004p166}. The full spectral, spatial and temporal information of the investigated spin wave modes
is now accessible.

\begin{figure}
\includegraphics[width=0.5\columnwidth]{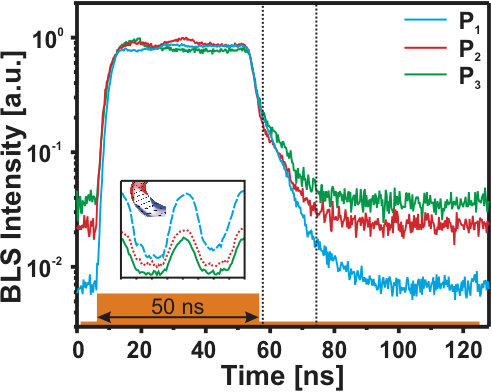}
\caption{\label{Fig_5} %
(color online) Temporal evolution of the quasi-eigenmodes located in the poles of the ring structure under the
resonant excitation with a microwave pulse with a duration of \mbox{50\,ns}. The upper curve (red) displays the
lowest frequency mode \mbox{$P_{3}$=2.5\,GHz} and the lowest curve (blue) corresponds to the quasi-eigenmode
\mbox{$P_{1}$=3.9\,GHz}. The intensity of all temporal profiles is normalized at the time where the excitation
pulse terminates. In between the vertical, dashed lines an exponential decay of the quasi-eigenmode amplitudes
can be observed with different decay constants for each mode. }
\end{figure}

\begin{figure}
\includegraphics[width=0.5\columnwidth]{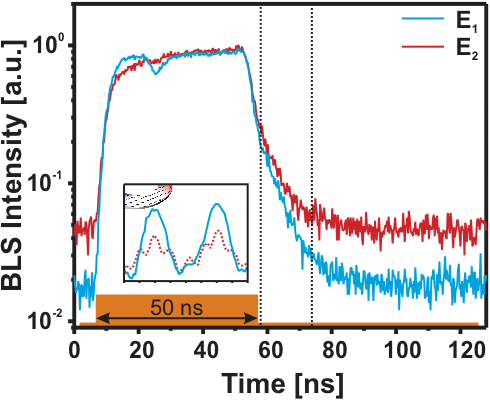}
\caption{\label{Fig_6}%
(color online) Temporal evolution of the quasi-eigenmodes located in the equator of the ring structure under
resonant excitation with a microwave pulse with a duration of \mbox{50\,ns}. The lower curve (blue) displays the
behavior of the quasi-eigenmode \mbox{$E_{1}$=7.3\,GHz}, the upper curve (red) corresponds to the quasi-eigenmode
\mbox{$E_{2}$=6.6\,GHz}. The intensity is normalized at the time where the excitation pulse terminates. The decay
constant $\tau$ is determined in the time window defined by the vertical, dashed lines. }
\end{figure}

Figure~\ref{Fig_5} shows the temporal profile of the spin-wave quasi-eigenmodes $P_{1}$, $P_{2}$, and $P_{3}$ in
the pole region of the onion state. The time in which the microwave pulse is applied is indicated in the bottom
of the figure. The spin-wave intensities increase drastically within the first few nanoseconds when the pulse
arrives and starts to decay when the pulse terminates. The intensities of all three measurements are normalized
to the intensity at the time when the excitation pulse terminates in order to emphasize the difference in the
relaxation part of the temporal profile. A multiplication with a constant factor does not change the slope of an
exponential decay in a logarithmic plot but it supports the comparison of the decay of the different spin-wave
quasi-eigenmodes. It can be seen in the time window between the two vertical dashed lines in Fig.\,\ref{Fig_5}
that the decay constant $\tau$, if one assumes an exponential decay $\exp({-\frac{t}{\tau}})$, differs for each
frequency. The higher the spin-wave frequency is the faster is the relaxation, i.e. the smaller is the decay
constant $\tau$. The same qualitative behaviour is observed for the spin waves in the equatorial region, see
Fig.\,\ref{Fig_6}. The first azimuthal mode with the higher frequency shows a faster relaxation than the third
azimuthal mode with lower frequency.

\begin{figure}
\includegraphics[width=0.5\columnwidth]{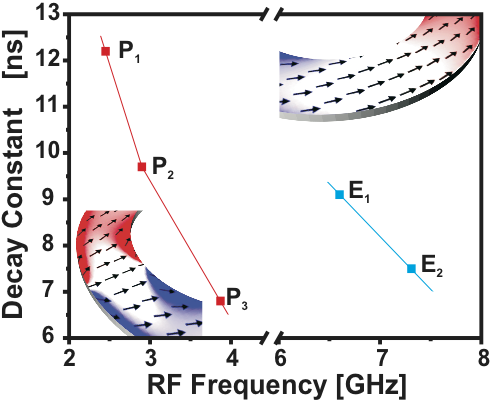}
\caption{\label{Fig_7}%
(color online) Decay constants $\tau$ for all investigated quasi-eigenmodes as a function of the excitation
frequency. In the left part of the figure $\tau$ is plotted for the low frequency spin waves confined in the pole
region \mbox{0$^{\circ}$} of the onion state. The right part shows $\tau$ of the spin waves at the equatorial
region \mbox{90$^{\circ}$}. The decay constant is extracted out of the temporal profiles in Figs.\,\ref{Fig_5}
and \ref{Fig_6} in the time windows indicated by the dashed vertical lines where the influence of the microwave
pulse is negligible. }
\end{figure}

The decay constants for all investigated spin-wave quasi-eigenmodes are summarized in Fig.\,\ref{Fig_7}. Here it
is evident that the decay constant is decreasing with increasing frequency for the same type of spin waves, but
there is a general difference when the character of the spin-wave quasi-eigenmode, i.e. the wavevector and the
quantization condition, is changed. We find - despite the fact that the frequencies of the spin waves are much
higher in the equatorial region - that the relaxation times are comparable to the relaxation times at the poles
of the onion state. Several mechanisms contribute to the dissipation of spin-wave quasi-eigenmodes, for example
coupling to the lattice, i.e. magnon-phonon-scattering, and coupling within the spin-wave quasi-eigenmode system,
i.e. magnon-magnon-scattering. The first mechanism depends on the magneto-elastic constants of the used material
and is, therefore, independent of the lateral position on the ring. The efficiency of energy transfer in the
latter mechanism strongly depends on the total internal field, the magnetization distribution within the
quantization volume and, since energy conservation has to be fulfilled in the magnon-magnon-scattering process,
on the frequencies of the participating spin waves. Hence, a position dependency of the dissipation processes, as
we observe it in the ring structure, is obvious and can explain that the decay constants $\tau$ are nearly the
same at the equator as in the poles - even if the frequencies of the measured spin waves are much higher.
Independent measurements show evidence for a direct coupling of spin wave modes excited near the poles of the
ring with spin waves confined to the equatorial region via three-magnon-scattering. This is evidenced by a
resonance in the coupling, i.e. maximum in the decay constant, when the frequency ratio between modes in the
equatorial and pole regions is 2:1 fulfilling energy conservation for three magnon scattering.

\section{Conclusions}
In conclusion we have determined the decay constants of all spin wave modes that can be excited with in-plane
microwave pulses in a small magnetic Ni$_{81}$Fe$_{19}$ ring using time-resolved BLS microscopy. Low frequency
excitations were observed in the poles of the onion state whereas the spin waves detected at the equatorial
position have higher frequencies and were identified as the well-known azimuthal modes of the onion state. The
decay constants were extracted from the temporal evolution of the spin wave amplitudes after the end of the
exciting microwave pulse. A decrease of the decay constant, i.e. a shorter lifetime of the spin-wave
quasi-eigenmodes, was found when the frequency of the quasi-eigenmode is higher. The comparison of the
dissipation time of the quasi-eigenmodes confined at the pole and the equatorial region of the onion state is
indicating that different dissipation mechanisms are responsible for the relaxation of the magnetization,
depending on the position and the quantization conditions of the spin-wave quasi-eigenmodes.

\section{Acknowledgment}
Support by the Priority Program SPP 1133 of the Deutsche Forschungsgemeinschaft and the NEDO International Joint
Research Program 2004IT093, Japan, is gratefully acknowledged. The authors acknowledge S. Trellenkamp and S.
Wolff from the Nano + Bio Center, University of Kaiserslautern, for technical support.

\end{document}